\documentclass[twocolumn,english,superscriptaddress,prl,preprintnumbers,amsmath,amssymb,floatfix,longbibliography]{revtex4-1}
\usepackage[T1]{fontenc}
\usepackage[utf8]{inputenc}
\usepackage{color}
\usepackage{babel}
\usepackage{amsmath}
\usepackage{graphicx}
\usepackage{esint}
\usepackage[unicode=true,pdfusetitle,
 bookmarks=true,bookmarksnumbered=false,bookmarksopen=false,
 breaklinks=false,pdfborder={0 0 0},pdfborderstyle={},backref=false,colorlinks=true]
 {hyperref}

\makeatletter
\usepackage{dcolumn}
\usepackage{bm}
\usepackage{color}
\usepackage{soul}
\usepackage{babel}
\usepackage{amsfonts}
\usepackage{slashed}
\usepackage{enumerate}
\usepackage{babel}
\usepackage{inputenc}

\makeatother

\begin{document}
\global\long\def\sgn{\mathrm{sgn}}%
\global\long\def\ket#1{\left|#1\right\rangle }%
\global\long\def\bra#1{\left\langle #1\right|}%
\global\long\def\sp#1#2{\langle#1|#2\rangle}%
\global\long\def\abs#1{\left|#1\right|}%
\global\long\def\avg#1{\langle#1\rangle}%

\title{Near-unit efficiency of chiral state conversion via hybrid-Liouvillian
dynamics}
\author{Parveen Kumar}
\affiliation{Department of Condensed Matter Physics, Weizmann Institute of Science,
Rehovot, Israel}
\author{Kyrylo Snizhko}
\affiliation{Institute for Quantum Materials and Technologies, Karlsruhe Institute
of Technology, 76021 Karlsruhe, Germany}
\affiliation{Department of Condensed Matter Physics, Weizmann Institute of Science,
Rehovot, Israel}
\author{Yuval Gefen}
\affiliation{Department of Condensed Matter Physics, Weizmann Institute of Science,
Rehovot, Israel}
\begin{abstract}
Following the evolution under a non-Hermitian Hamiltonian (nHH) involves
significant probability loss. This makes various nHH effects impractical
in the quantum realm. In contrast, Lindbladian evolution conserves
probability, facilitating observation and application of exotic effects
characteristic of open quantum systems. Here we are concerned with
the effect of chiral state conversion: encircling an exceptional point,
multiple system states are converted into a single system eigenmode.
While for nHH the possible converted-into eigenmodes are pure states,
for Lindbladians these are typically mixed states. We consider hybrid-Liouvillian
evolution, which interpolates between a Lindbladian and a nHH and
enables combining the best of the two worlds. We design adiabatic
evolution protocols that give rise to chiral state conversion with
\emph{pure} final states, no probability loss, and high fidelity.
Furthermore, extending beyond continuous adiabatic evolution, we design
a protocol that facilitates conversion to pure states with fidelity
1 and, at the same time, no probability loss. Employing recently developed
experimental techniques, our proposal can be implemented with superconducting
qubit platforms.
\end{abstract}
\maketitle
\emph{Introduction.---}Over the past few years, exceptional points
(EPs) of non-Hermitian Hamiltonians \citep{Kato1995,Berry2004a,Heiss2012,Ashida2020}
have become an object of intense study in a number of distinct contexts.
From rather abstract studies involving $\mathcal{PT}$-symmetric nHHs
\citep{Bender1998,Bender2007,Jiankeeyang2018,Bender2019}, adiabatic
nHH evolution \citep{Mailybaev2005,Berry2011,Milburn2015,Doppler2016,Hassan2017,Pick2017,Zhang2018,Zhong2018,Miri2019,Liu2020h},
topological manifestations~\citep{Wang2009,Lefebvre2009,Uzdin2011,Rechtsman2013,Chang2014,El-Ganainy2018},
exotic band structures \citep{Zhen2015}, and properties of non-equilibrium
phase transitions \citep{Altland2020,Soriente2021} to more practically-oriented
proposals of loss-induced transparency and lasing~\citep{Guo2009,Peng2014a,Jing2014,Brandstetter2014,Miao2017,Zhang2018a},
on-demand directional emission~\citep{Kim2014,Wiersig2014a,Peng2016,Li2020p},
optimal energy transfer~\citep{Xu2016,Assawaworrarit2017}, and enhanced
sensing~\citep{Chen2017,Hodaei2017,Zhang2019b}.

One of the most striking effects emerging in the vicinity of an EP
is chiral state conversion \citep{Mailybaev2005,Berry2011,Milburn2015,Doppler2016,Hassan2017,Pick2017,Zhang2018,Zhong2018,Miri2019}:
Under adiabatic evolution of the system along a trajectory in the
parameter space such that an EP is encircled, all possible initial
states of the system are converted to one final state. The final state
corresponds to one of the system's eigenmodes. The directionality
of the winding around the EP (the chirality) determines which eigenmode
it is. A major hindrance in employing this chiral effect for practical
purposes or incorporating it in more complex manipulation protocols,
are the significant losses incurred under nHH evolution. These may
be acceptable in \emph{classical} applications; in the \emph{quantum}
context, any loss of probability is highly detrimental.

While classical lossy systems are often naturally described in terms
of a nHH, open quantum systems only allow for such a description in
the presence of postselection, based on monitoring of the environment
\citep{Naghiloo2019,Chen2021}. In the presence of a non-monitored
Markovian environment, quantum system evolution is described by the
Lindblad master equation~\citep{Carmichael1999,Gardiner2004,Haroche2006,Breuer2007}.
The physics of EPs of Lindbladian superoperators is attracting much
attention \citep{Hatano2019b,Pick2019,Minganti2019,Minganti2020b,Arkhipov2020a,Arkhipov2020b,Gunderson2020,Kumar2021,Gulacsi2021,Khandelwal2021},
in particular, in the context of optimal state preparation and stabilization
\citep{Pick2019,Kumar2021,Khandelwal2021}. Lindbladian evolution
conserves the total probability, eliminating losses and, naively,
opening the way to efficient chiral state conversion in the quantum
realm. However, generically the eigenmodes involved (i.e., the potential
conversion results) do not correspond to pure states, limiting the
applicability of such protocols \citep{Pick2019}.

\begin{figure}
	\begin{centering}
		\includegraphics[width=1\columnwidth]{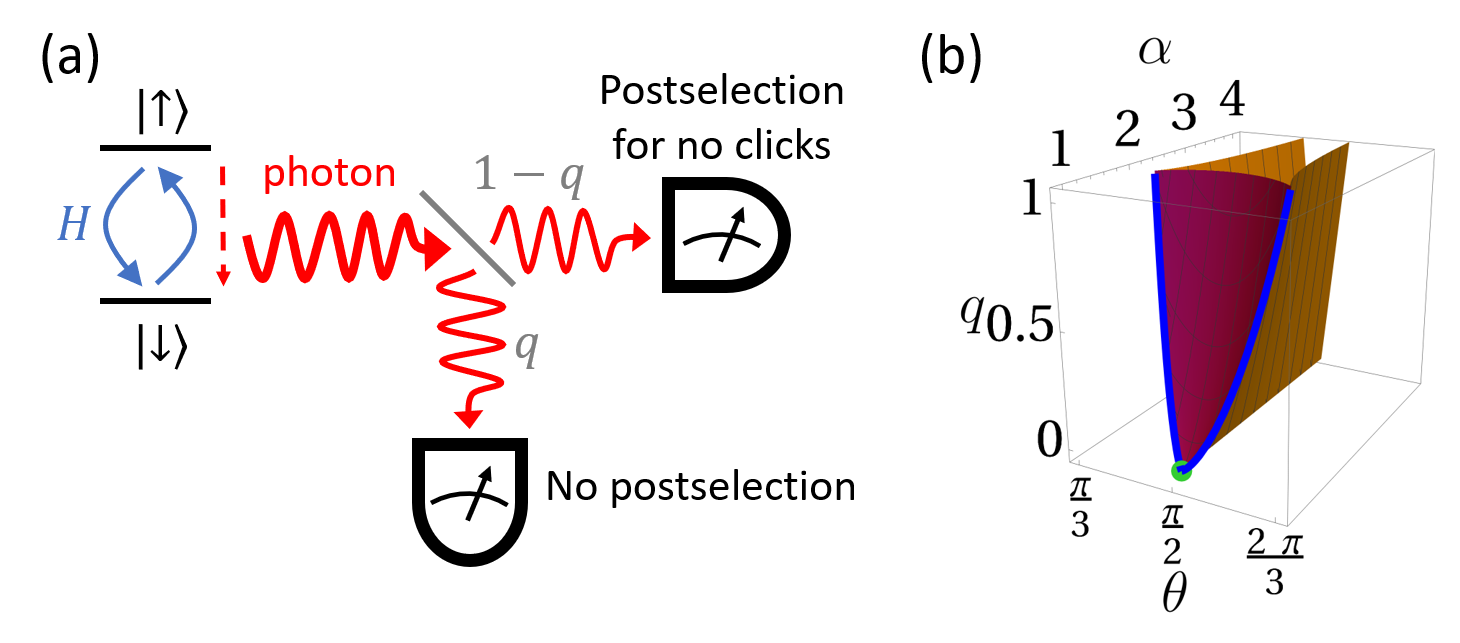}
		\par\end{centering}
	\caption{\label{fig:setup+EPsurface} (a)---The setup under consideration.
		A qubit is subject to a Hamiltonian evolution and relaxation processes.
		Whenever a relaxation event takes place, the experimental run is discarded
		with a prescribed probability, $1-q$. (b)---The manifold of EPs
		of the hybrid-Liouvillian $\mathcal{L}_{q}$, cf.~Eqs.~(\ref{eq:system-master-equation}-\ref{eq:hybrid-Liouvillian}),
		in the parameter space of $\theta$ (specifying the Hamiltonian),
		$q$ and $\alpha=\gamma/2\omega$ (determining the relative strength
		of the Hamiltonian and relaxation). The green point ($\alpha=1,~\theta=\frac{\pi}{2},~q=0$) corresponds to
		a 4th order degeneracy (featuring, however, only a 3rd order EP) ---
		the hybrid-Liouvillian counterpart of the EP of the nHH $\tilde{H}$,
		cf.~Eq.~(\ref{eq:system-master-equation-without-jumps}). At $q>0$
		the degenerate manifold splits into two lines (blue) of 3rd order
		EPs connected by a surface of 2nd order EPs (purple). Two more surfaces
		of 2nd order EPs (orange) are bounded by the blue lines and continue
		up to $\alpha\rightarrow\infty$; their intersection at $q=0$ forms
		a line of 2nd order degeneracies, which are not EPs.}
\end{figure}

Here we investigate chiral state conversion in a system featuring
a controllable degree of postselection, cf.~Fig.~\ref{fig:setup+EPsurface}(a).
The dynamics is described within the hybrid-Liouvillian (hL) formalism
\citep{Minganti2020b}. Depending on the postselection parameter,
$q$, the hL interpolates between nHH ($q=0$) and Lindbladian ($q=1$)
dynamics. We show that the EPs of the hL form a rich structure within
the parameter space extended by $q$. This structure continuously
connects the EPs of the Lindbladian at $q=1$ to those of the corresponding
nHH at $q=0$, opening the way for chiral state conversion with the
degree of postselection being varied during the protocol. This way
the best of the two worlds can be combined: low losses, inherent to
$q\approx1$ evolution, and pure state eigenmodes, inherent to $q=0$.

Below we show that varying $q$ during the evolution alongside the
other parameters significantly reduces the probability loss, while
the mode conversion fidelity (that quantifies the accuracy of mode
conversion) remains almost unchanged. We note, though, that even with
the protocol where $q$ is varied during the adiabatic evolution,
the two desired goals: (i) perfect values of unit fidelity and (ii)
no loss due to postselection, are unattainable simultaneously. To
achieve these two goals together, we further employ a ``hopping strategy'',
developed recently in the context of nHH dynamics~\citep{Li2020p}:
we replace parts of the adiabatic encircling trajectory by abrupt
hops. We demonstrate that combining the hopping strategy with a controlled
postselection parameter, $q$, facilitates achieving both goals mentioned
above, cf. Fig.~\ref{fig:conversion_performance}.

\emph{Model.---}We study a single qubit subject to a Hamiltonian
evolution and spontaneous relaxation processes, cf. Fig.~\ref{fig:setup+EPsurface}(a).
When the qubit is in an excited state, $\ket{\uparrow}$, it can relax
to the ground state, $\ket{\downarrow}$, by emitting a photon. Whenever
a photon is emitted, the experimental run is either interrupted
and discarded from the statistics (probability $1-q$) or is allowed to proceed (probability $q$) depending on which detector clicks \citep{ImperfectDetectorFoot}. The resulting evolution of the qubit density matrix is described
by the following Liouville equation:
\begin{equation}
\frac{d\rho}{dt}=\mathcal{L}_{q}[\rho]=-i[H,\rho]-\frac{\gamma}{2}\left(\{L^{\dagger}L,\rho\}-2qL\rho L^{\dagger}\right),\label{eq:system-master-equation}
\end{equation}
with the Hamiltonian
\begin{equation}
H=\frac{\omega}{2}\left(\sin\theta\,\sigma_{x}+\cos\theta\,\sigma_{z}\right)\equiv\frac{\omega_{x}}{2}\sigma_{x}+\frac{\omega_{z}}{2}\sigma_{z},\label{eq:system-Hamiltonian}
\end{equation}
the Lindblad jump operator $L=\ket{\downarrow}\bra{\uparrow}$, and
the relaxation rate $\gamma$. Note that Eq.~(\ref{eq:system-master-equation})
does not preserve the probability (density matrix trace) unless $q=1$,
so it is not a proper Liouvillian evolution (hence the name of hybrid-Liouvillian
\citep{Minganti2020b}).

In the absence of postselection, $q=1$, one recovers the standard
Lindblad equation \citep{Carmichael1999,Gardiner2004,Haroche2006,Breuer2007}.
In the case of perfect postselection, $q=0$, the system evolution
can be described in terms of a non-Hermitian Hamiltonian $\tilde{H}$:
\begin{equation}
\frac{d\rho}{dt}=-i\left(\tilde{H}\cdot\rho-\rho\cdot\tilde{H}^{\dagger}\right),\quad\tilde{H}=H-i\frac{\gamma}{2}\,L^{\dagger}L.\label{eq:system-master-equation-without-jumps}
\end{equation}

We consider the range of parameters to be $\omega\geq0$ and $\theta\in[0,\pi]$.
We further define a dimensionless parameter $\alpha=\frac{\gamma}{2\omega}\geq0$.

\emph{Hybrid-Liouvillian Exceptional Points.---}To model the chiral
behavior marking the hL dynamics of encircling EPs in parameter space,
first we need to investigate the EPs of the superoperator $\mathcal{L}_{q}$
defined in Eq.~(\ref{eq:system-master-equation}). The present, rather
technical section, summarizes the steps taken to obtain Fig.~\ref{fig:setup+EPsurface}(b).
For that purpose, it is convenient to write the matrix representation
of the superoperator:
\begin{equation}
\mathcal{L}_{q}=\left(\begin{array}{cccc}
-\gamma & i\frac{\omega_{x}}{2} & -i\frac{\omega_{x}}{2} & 0\\
i\frac{\omega_{x}}{2} & -\frac{\gamma}{2}-i\omega_{z} & 0 & -i\frac{\omega_{x}}{2}\\
-i\frac{\omega_{x}}{2} & 0 & -\frac{\gamma}{2}+i\omega_{z} & i\frac{\omega_{x}}{2}\\
\gamma q & -i\frac{\omega_{x}}{2} & i\frac{\omega_{x}}{2} & 0
\end{array}\right)\label{eq:hybrid-Liouvillian}
\end{equation}
in the basis $\{\rho_{\uparrow\uparrow},\rho_{\uparrow\downarrow},\rho_{\downarrow\uparrow},\rho_{\downarrow\downarrow}\}$.
The superoperator's eigenvalues $\{\lambda\}$ correspond to zeros
of the characteristic polynomial $\mathcal{C}_q(\lambda)=\text{Det}(\mathcal{L}_{q}-\lambda\mathbb{\,I})$,
where $\mathbb{I}$ is the identity matrix. An $n$th order EP corresponds
to an $n$th order degeneracy where $n$ eigenvectors coalesce into
a single eigenvector. We first look for the set of $n$th order degeneracies
by requiring $\mathcal{C}_q(\lambda_{d})=...=\mathcal{C}_q^{(n-1)}(\lambda_{d})=0$,
where $\mathcal{C}_q^{(k)}(\lambda)$ denotes the $k$th derivative
of the characteristic polynomial. We then check the number of linearly
independent eigenvectors corresponding to this $\lambda_{d}$ in order
to separate EPs from trivial degeneracies.

Figure \ref{fig:setup+EPsurface}(b) shows the resulting locations
of the EPs and non-EP degeneracies in the space of protocol parameters
$(\alpha,\theta,q)$ \citep{SupplMat}. We find a 4th order degeneracy
located at $(1,\pi/2,0)$ (green point). At $q=0$, the system can
be described by a nHH, $\tilde{H}$ in Eq.~(\ref{eq:system-master-equation-without-jumps}),
which has a well-studied \citep{Milburn2015,Doppler2016,Hassan2017,Zhang2018,Miri2019,Li2020p}
2nd order EP at $\alpha=1$, $\theta=\pi/2$. In the language of hL,
this nHH EP becomes a 4th order degeneracy. However, it is only a
3rd order EP: only three eigenvectors of hL coalesce while the fourth
remains separate. At $(\alpha>1,\pi/2,0)$, we also find a line of
2nd order degeneracies of $\mathcal{L}_{q}$, which are not EPs.

As soon as $q\neq0$ we find a more involved spectrum of EPs. At each
$q>0$, we find two EPs of 3rd order and three lines of 2nd order
EPs. Taken for all $q\in(0;1]$ the structure becomes two lines of
3rd order EPs and three surfaces of 2nd order EPs, cf.~Fig.~\ref{fig:setup+EPsurface}(b).

\begin{figure}
\begin{centering}
\includegraphics[width=1\columnwidth]{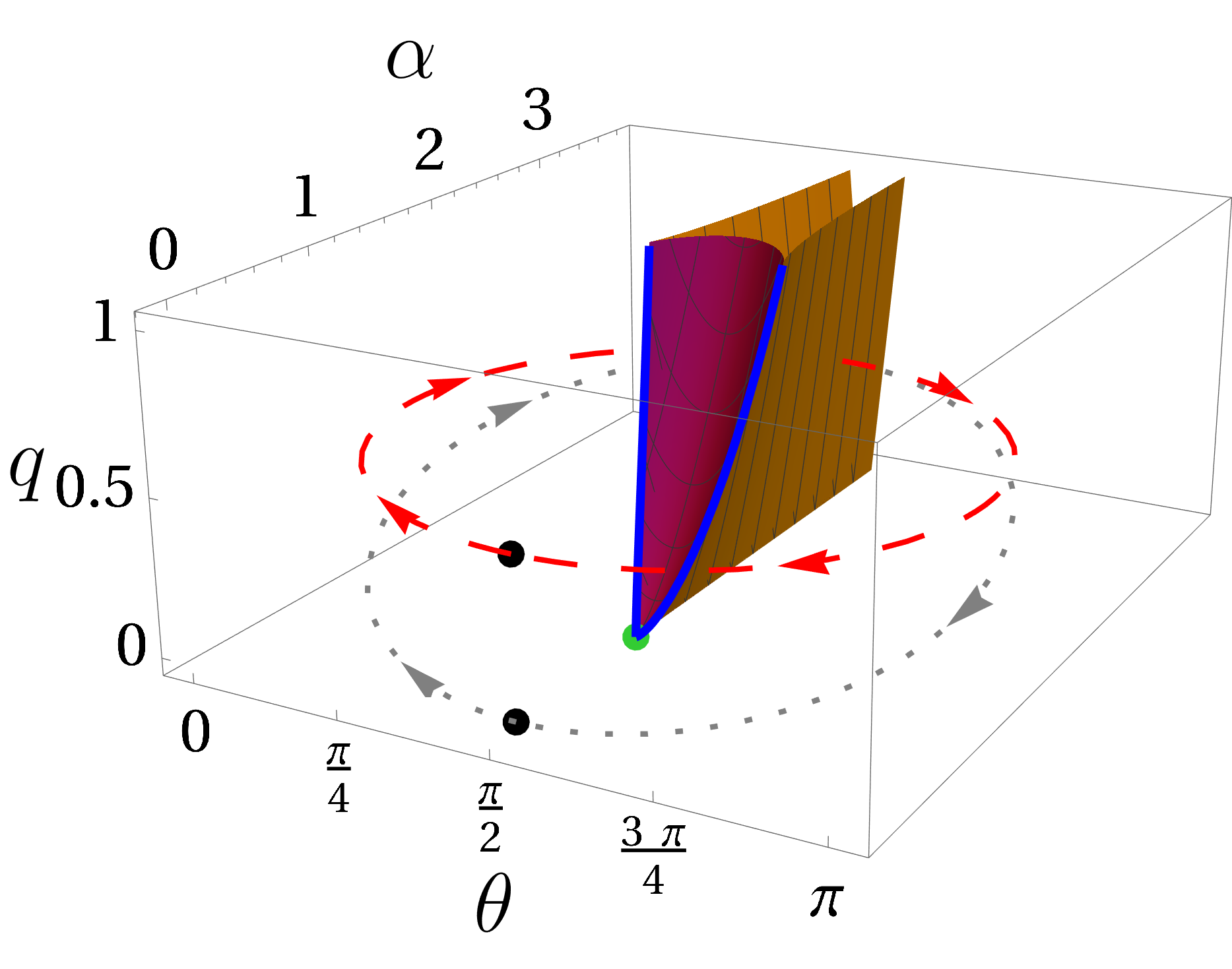}
\par\end{centering}
\caption{\label{fig:trajectories_slow} Trajectories encircling the central
part of the EP structure. The tilted trajectories ($q\protect\ne\text{const}$,
dotted gray), defined in Eq.~(\ref{eq:tilted_trajectory}), all start
at the same point $(\alpha,\theta,q)=(0,\pi/2,0)$, but can approach
different values of $q_{0}$ in the middle of the closed trajectory.
Flat trajectories (dashed red), defined in Eq.~(\ref{eq:flat_trajectory}),
correspond to $q=q_{0}$ plane. The black dots represent the starting
(and end) point of the corresponding trajectories.}
\end{figure}

\emph{X-adiabatic encircling of the EP structure.---}We are now in
a position to discuss chiral state conversion in the system. We start
with some abstract arguments for why it could be expected and what
the caveats are. We then resort to a numerical investigation and discuss
its results confirming and quantifying the expected behavior.

Consider the EP manifold in the $(\alpha,\theta,q)$ space, cf.~Fig.~\ref{fig:setup+EPsurface}(b).
Let us for a moment ignore the fact that the EP manifold extends to
$\alpha\rightarrow\infty$ and pretend that one can encircle it. The
\emph{chirality} of chiral state conversion originates in the switching
of the system eigenmodes as one goes around an exceptional point \citep{Zhong2018}.
The switching itself stems from the spectrum non-analiticity at the
EP. Since the EP manifold is continuous as a function of $q$, the
effective non-analiticity of the entire EP manifold should be the
same at any given $q$. This implies that the switching of the respective
eigenmodes at different $q$ obeys the same rules. Therefore, one
expects the same chiral state conversion behavior at all $q$. Note,
however, that the eigenmodes of the hL are $q$-dependent, i.e., the
final state may depend on the value of $q$ at the beginning/end of
the trajectory.

Recall now that the structure extends to $\alpha\rightarrow\infty$,
so that encircling the entire structure is impossible. It follows
that it is impossible to encircle even part of the structure in a
completely adiabatic way. For example, trying to encircle the lines
of 3rd order EPs one must cross the surfaces of 2nd order EPs, cf.~Fig.~\ref{fig:trajectories_slow}.
On that sub-manifold some of the eigenvalues coincide making it impossible
to satisfy the adiabaticity condition \citep{AdiabaticityFoot}. We call such evolution x-adiabatic, i.e. adiabatic everywhere except
for a few points along the trajectory.

In order to investigate chiral state conversion under x-adiabatic
evolution, we perform a numerical investigation for two families of
closed trajectories, cf.~Fig.~\ref{fig:trajectories_slow}. The
first family comprises trajectories that start and end in the $q=0$
plane, with the postselection rate varied along the trajectory. The
second family comprises trajectories with a fixed $q$, i.e. the postselection
rate remains constant along the trajectory. In reference to their
shape, we denote the first family \emph{tilted} and the second one
\emph{flat}. The tilted trajectories start at $(0,\pi/2,0)$, then
reach $(\alpha_{0},\pi/2,q_{0})$ at the mid-point, and finally return
back to the initial point:
\begin{gather}
\alpha_{\textrm{tilted}}(t)=3\sin^{2}\frac{\pi t}{T},\quad q_{\textrm{tilted}}(t)=q_{0}\sin^{2}\frac{\pi t}{T},\label{eq:tilted_trajectory}\\
\theta_{\textrm{tilted}}(t)=\frac{\pi}{2}-\frac{3}{2}\sin\frac{2\pi\chi t}{T},\nonumber
\end{gather}
where the evolution takes place within the time interval $t\in[0,T]$,
and $\chi=\pm1$ corresponds to different winding chiralities. The
flat trajectories start/end at $(0,\pi/2,q_{0})$ and remain in $q=q_{0}$
plane at all intermediate times:
\begin{gather}
\alpha_{\textrm{flat}}(t)=3\sin^{2}\frac{\pi t}{T},\quad q_{\textrm{flat}}(t)=q_{0},\label{eq:flat_trajectory}\\
\theta_{\textrm{flat}}(t)=\frac{\pi}{2}-\frac{3}{2}\sin\frac{2\pi\chi t}{T}.\nonumber
\end{gather}

Consider first the tilted trajectories. The $q_{0}=0$ trajectory
can be equivalently described by a nHH $\tilde{H}$, cf.~Eq.~(\ref{eq:system-master-equation-without-jumps}).
This problem is well studied \citep{Milburn2015,Doppler2016,Hassan2017,Zhang2018,Miri2019,Li2020p}.
At the initial and final point, the system experiences only the Hamiltonian
$H$, cf.~Eq.~(\ref{eq:system-Hamiltonian}), whose eigenstates
are $\ket{\pm}=(\ket{\uparrow}\pm\ket{\downarrow})/\sqrt{2}$. Adiabatically
($T\rightarrow\infty$) following this trajectory at $q_{0}=0$ in
the clockwise direction ($\chi=+1$) leads to a conversion of any
initial system state to $\rho_{\chi=+1}=\ket +\bra +$. Following
the trajectory in the opposite direction ($\chi=-1$) converts any
initial state to $\rho_{\chi=-1}=\ket -\bra -$. Note that in the
hL language this trajectory is x-adiabatic (the trajectory crosses
the line of 2nd order degeneracies at $(\alpha>1,\theta=\pi/2,q=0)$).
Nevertheless, the outcome of the evolution should coincide with the
prediction of the nHH formalism.

For $q_{0}\neq0,$ we observe the same conversion behavior: any initial
state is converted into $\rho_{\chi}$ corresponding to the respective
direction $\chi$, as quantified by the fidelity $F=\mathrm{Tr}\,\rho(T)\rho_{\chi}$,
cf.~Fig.~\ref{fig:conversion_performance}(a). Note that the conversion
fidelity for tilted trajectories is almost independent of the value
of $q_{0}$.

The protocol, however, comes with a significant probability loss.
The dependence of the probability $P=\mathrm{Tr}\,\rho(T)$ of carrying
out the experiment to the end, without having to discard it due to
postselection, is presented in Fig.~\ref{fig:conversion_performance}(b).
As expected, $P$ increases with $q_{0}$: for higher $q_{0}$ less
postselection is applied throughout the trajectory. Yet, even with
$q_{0}=1$, $P\approx10^{-2}$, far from being practically useful.

\begin{figure}
\begin{centering}
\includegraphics[width=1\columnwidth]{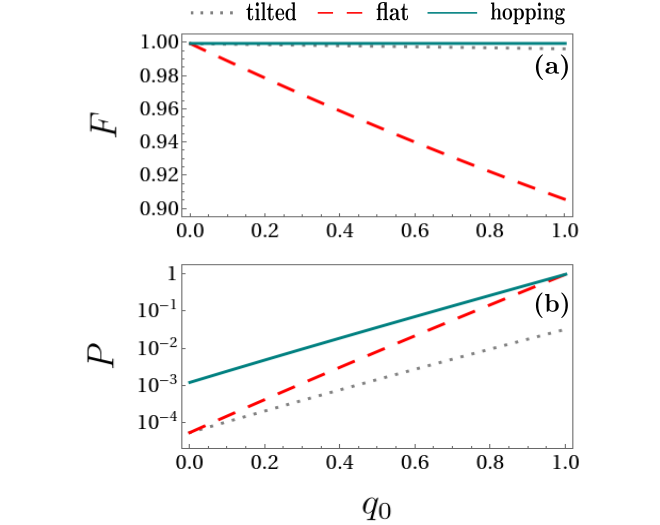}
\par\end{centering}
\caption{\label{fig:conversion_performance}State conversion fidelity $F=\mathrm{Tr}\,\rho(T)\rho_{\chi}$
and postselection probability $P=\mathrm{Tr}\,\rho(T)$ for tilted
(dotted gray, Eq.~(\ref{eq:tilted_trajectory})), flat (dashed red,
Eq.~(\ref{eq:flat_trajectory})), and hopping (solid teal, Eqs.~(\ref{eq:hopping_trajectory_first}-\ref{eq:hopping_trajectory_last}))
trajectory as a function of the postselection parameter $q_{0}$.
For all trajectories we took a maximally mixed initial state, $\rho_{i}=\frac{1}{2}\protect\ket +\protect\bra ++\frac{1}{2}\protect\ket -\protect\bra -$,
$\omega=1~ns^{-1}$ for the Liouvillian $\mathcal{L}_{q}$ (cf.~Eqs.~(\ref{eq:system-master-equation}--\ref{eq:hybrid-Liouvillian}))
and a total evolution time $T=100 ~ns$. For the hopping trajectory, $T_{1}=0.2T$,
$T_{2}=0.6T$, $\alpha_{\mathrm{i}}=10^{-5}$, and $\alpha_{\mathrm{ii}}=10$.
The values of $F$ and $P$ for clockwise ($\chi=+1$) and counterclockwise
($\chi=-1$) trajectories coincide, hence only three curves are shown.}
\end{figure}

For flat trajectories, the postselection probability can be much higher,
cf.~Fig.~\ref{fig:conversion_performance}(b). In particular, for
$q_{0}=1$ there is no probability loss (as no postselection is applied).
However, this comes at the price of a significant fidelity loss. Let
us briefly explain the reasons for the latter. The initial/final point
of the flat trajectory is $(\alpha,\theta,q)=(0,\pi/2,q_{0})$, so
that $\gamma=2\omega\alpha=0$. Therefore, the system eigenmodes at
the initial/final point of the trajectory are determined by the same
parameters of $H$ in Eq.~(\ref{eq:system-Hamiltonian}), independintly
of $q_{0}$. However, as soon as $\alpha\neq0$ the system's eigenmodes
do depend on $q_{0}$. While for the system governed by a nHH ($q_{0}=0$)
all system eigenmodes (including the one to which all the others are
converted) are pure states, for a system governed by a hL these are
mixed states. Yet, if one aims to obtain the \emph{pure} state $\rho_{\chi}$,
this is achieved with fidelity $F\gtrsim0.9$.

\emph{Chiral state conversion with near-unit efficiency.---}
Recently, Ref.~\citep{Li2020p} proposed employing the strategy of parameter hopping for nHH and was able to achieve chiral state conversion with $P \approx F \approx 0.9$. While this is an impressive improvement over adiabatic protocols, it is still not good enough for properly quantum applications. Here
we design a protocol in which both the fidelity $F$ and the postselection
probability $P$ can simultaneously have values approaching $1$ with
arbitrarily high accuracy. This is achieved by combining variable postselection rate $q$ with parameter hopping.

\begin{figure}
\begin{centering}
\includegraphics[width=1\columnwidth]{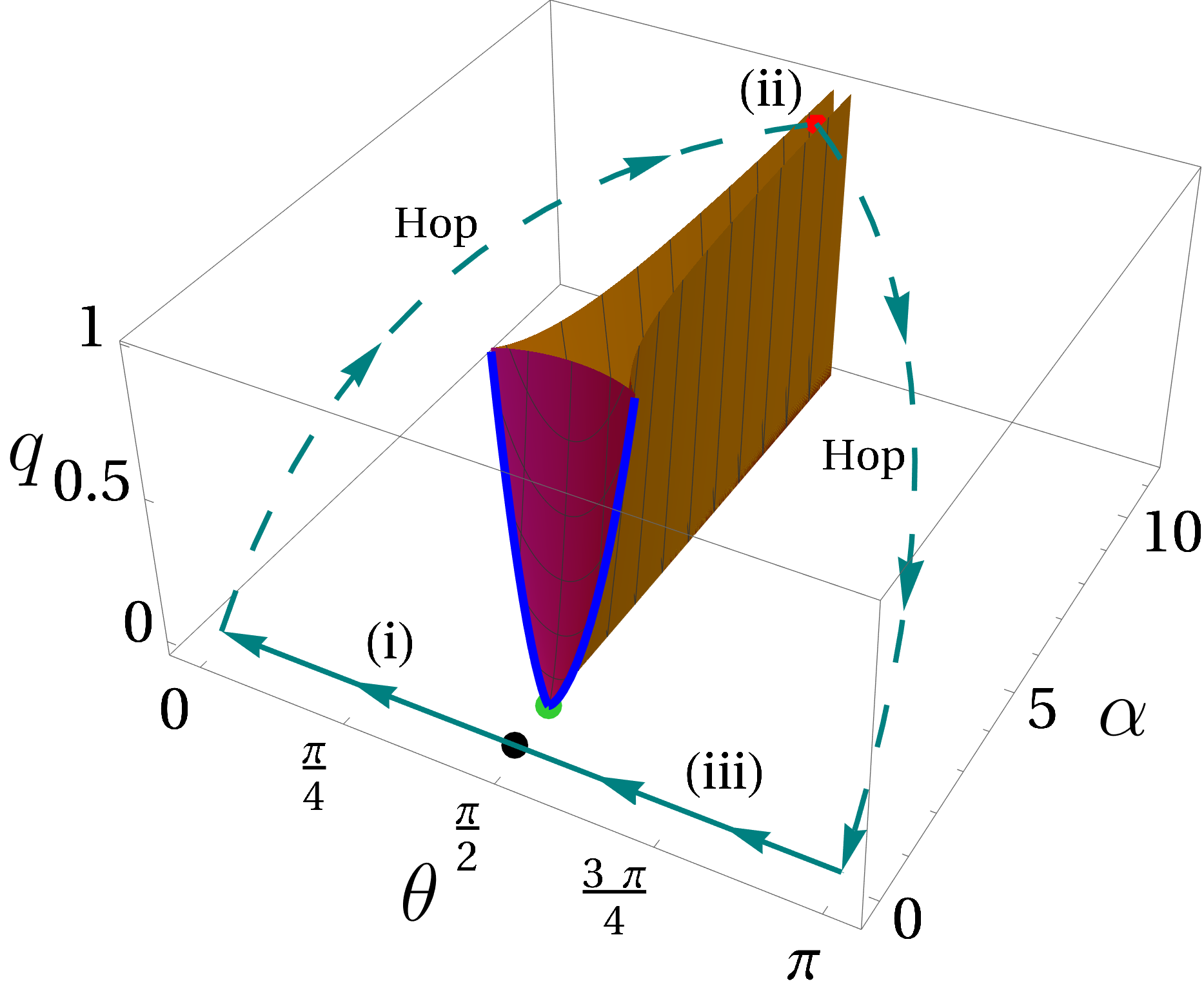}
\par\end{centering}
\caption{\label{fig:trajectory_hopping} The hopping trajectory with $q_{0}=1$
and $\chi=+1$, cf.~Eqs.~(\ref{eq:hopping_trajectory_first}--\ref{eq:hopping_trajectory_last}),
involves continuous adiabatic evolution at $q=0$ {[}(i), (iii){]}
and a relaxation-dominated evolution at $q=1$ (ii), as well as and
the instantaneous hops between them. This minimizes the probability
loss, while optimizing the conversion fidelity. The black dot shows
the initial (and final) point of the trajectory. The entire part (ii)
corresponds to the red dot ($\alpha=10,~\theta=\frac{\pi}{2},~ q=1$), cf.~Eq.~(\ref{eq:hopping_trajectory}).}
\end{figure}

Consider the following trajectory in parameter space:

\begin{multline}
(\mathrm{i})\quad t\in[0,T_{1}):\alpha_{\textrm{hopping}}(t)=\alpha_{\mathrm{i}}\rightarrow0,\\
\theta_{\textrm{hopping}}(t)=\frac{\pi}{2}\left(1-\chi\frac{t}{T_{1}}\right),\;q_{\textrm{hopping}}(t)=0;\label{eq:hopping_trajectory_first}
\end{multline}

\begin{multline}
(\mathrm{ii})\quad t\in(T_{1},T_{1}+T_{2}):\alpha_{\textrm{hopping}}(t)=\alpha_{\mathrm{ii}}\rightarrow\infty,\\
\theta_{\textrm{hopping}}(t)=\frac{\pi}{2},\;q_{\textrm{hopping}}(t)=q_{0};\label{eq:hopping_trajectory}
\end{multline}

\begin{multline}
(\mathrm{iii})\quad t\in(T_{1}+T_{2},T=2T_{1}+T_{2}):\alpha_{\textrm{hopping}}(t)=\alpha_{\mathrm{i}},\\
\theta_{\textrm{hopping}}(t)=\frac{\pi}{2}\left(1-\chi\frac{t-T}{T_{1}}\right),\;q_{\textrm{hopping}}(t)=0.\label{eq:hopping_trajectory_last}
\end{multline}
This trajectory is depicted in Fig.~\ref{fig:trajectory_hopping}.
Note the discontinuous hops at times $T_{1}$ and $T_{1}+T_{2}$.
This trajectory enables chiral state conversion through the following
mechanism. For the trajectory's initial point, $(0,\pi/2,0)$, the
system's evolution is goverened by the Hamiltonian $H$ whose eigenstates
are $\ket{\pm}$. During (i) the system evolves adiabatically for
time $T_{1}$ so that $\ket +$ and $\ket -$ are transformed into
$\ket{\uparrow}$ and $\ket{\downarrow}$ respectively (for $\chi=+1$).
During (ii) the system stays at a single point in the parameter space,
and its dynamics is dominated by relaxation, so that the system eventually
ends up in $\ket{\downarrow}$. During (iii) the system is again governed
by $H$, so that $\ket{\downarrow}$ is adiabatically transformed
into $|+\rangle$. For $\chi=-1$ the roles of $\ket +$ and $\ket -$
are interchanged.

The dependence of the conversion fidelity $F$ and postselection probability
$P$ on $q_{0}$ is shown in Fig.~\ref{fig:conversion_performance}.
With the parameters used in the numerical simulation, $F>0.999$ for
all $q_{0}$ and $P>0.999$ when $q_{0}=1$.

In principle, $F=1$ and $P=1$ can be achieved. For this, one needs:
(a) $q_{0}=1$ and $\alpha_{\mathrm{i}}=0$, so that no part of the
trajectory involves probability losses, (b) $\alpha_{\mathrm{ii}}\rightarrow\infty$
and $T_{2}\rightarrow\infty$ for perfect state conversion in step
(ii), and (c) $T_{1}\rightarrow\infty$ for perfectly adiabatic transfer
in steps (i) and (iii). We point out that the locations of hops
need to be chosen carefully in order for unit efficiency to be possible.

\emph{Summary and discussion.---}We have investigated the chiral
state conversion under hybrid-Liouvillian dynamics. We have shown
that the effect can take place in such a setting and have designed
a protocol for chiral state conversion with \emph{pure target states}
and \emph{no probability loss}.

Designing this protocol is facilitated by the fact that the EPs of
a hL form a structure that continuously connects the EPs of a Lindbladian
and the respective nHH. This continuity implies that the chiral state
conversion effect, which is known for nHH and, separately, for Lindbladians
persists even when the degree of postselection is varied during the
protocol.

Our findings regarding the continuity of the EP structure
are applicable to generic single-, few-, and many-body systems governed
by hL. This follows from the continuous dependence of the superoperator $\mathcal{L}_{q}$ and its characteristic equation $\mathcal{C}_{q}(\lambda)$ on $q$. The EP structure continuity is a main ingredient for our near-unit-efficiency
chiral state conversion protocol. Therefore, it appears likely that our protocol
can be generalized to other systems as well.

With efficient many-body chiral state conversion,
one might envision applying it for quantum-annealing-like computations.
That is, encircling an EP structure in order to convert the system to
a state that represents a solution for some problem.

The setup considered here is closely related to recent experiments
investigating EPs in a superconducting qubit experiencing nHH \citep{Naghiloo2019}
and Lindbladian \citep{Chen2021} dynamics. \emph{Continuous} variation of the postselection parameter, needed for our x-adiabatic protocols, can only be done in the range $q=0.8-1$, as current efficiency of single-photon detection, which is needed for postselection, is $\lesssim 20\%$ \citep{Minev2019}. However, our \emph{hopping} protocol does not suffer from this restriction and is amenable to experimental test by discontinuous switching between nHH and Lindbladian evolutions in the course of experiment.

\begin{acknowledgments}
We thank Adi Pick for useful discussions. The authors acknowledge
the following financial support: the Deutsche Forschungsgemeinschaft
(DFG, German Research Foundation) - Projektnummer 277101999 -TRR 183
(project C01) (PK, KS, YG) , GO 1405/6-1 (KS), EG 96/13-1 (PK and
YG), the Israel Science Foundation (ISF) (YG), the National Science
Foundation through award DMR- 2037654 and the US-Israel Binational
Science Foundation (BSF) (YG), Jerusalem, Israel, and the Helmholtz
International Fellow Award (YG).

P.K. and K.S. contributed equally to this work.
\end{acknowledgments}

\bibliography{extra,exceptional_points_in_the_Hybrid-Liouvillian_formalism}

\begin{thebibliography}{62}
\bibitem [{Imp()}]{ImperfectDetectorFoot}%
  \BibitemOpen
  \href@noop {} {}\bibinfo {note} {We assume here perfect detector efficiency.
  Detector inefficiency can be incorporated into the value of $q$. Conversely,
  such a protocol can be viewed as a simulation of a finite-efficiency
  detector.}\BibitemShut {Stop}%
\end{thebibliography}

%\appendix
\clearpage
\renewcommand{\thefigure}{S\arabic{figure}}
\renewcommand{\figurename}{Supplemental Figure}
\setcounter{figure}{0}
\onecolumngrid

\section*{\large SUPPLEMENTARY INFORMATION}
%\part*{Supplemental material}
\section{The analytical formulas for the locations of the hybrid-Liouvillian
	EPs}

Here we present the analytical formulas for the exceptional points
(EPs) of the hybrid Liouvillian $\mathcal{L}_{q}$ given in Eq.~(\ref{eq:hybrid-Liouvillian})
of the main text.

\subsection{Fourth-order degeneracies\label{subsec:Fourth-order-EP}}

A fourth-order degeneracy implies that the characteristic polynomial,
$\mathcal{C}_q(\lambda)=\text{Det}(\mathcal{L}_{q}-\lambda\mathbb{\,I})$,
should satisfy the following constraints: $\mathcal{C}_q(\lambda)=0,\,\mathcal{C}_q^{\prime}(\lambda)=0,\,\mathcal{C}_q^{\prime\prime}(\lambda)=0,\,\,\text{and}\,\,\mathcal{C}_q^{\prime\prime\prime}(\lambda)=0$
where $^{\prime}$ denotes a derivative with respect to $\lambda$.
Solving these constraints simultaneously, and denoting $\alpha=\gamma/2\omega$,
we observe that there exists only one fourth-order degeneracy that
is located at

\begin{equation}
\alpha=1,\,\theta=\frac{\pi}{2},\,q=0.\label{eq:4-deg-conditions}
\end{equation}
The Jordan decomposition of $\mathcal{L}_{q}$ at these parameters
shows that this is a 3rd order EP with the fourth eigenvalue accidentally
coinciding with the other three.

\subsection{Third-order degeneracies\label{subsec:Third-order-EPs}}

In this case, the characteristic polynomial should satisfy the following
constraints: $\mathcal{C}_q(\lambda)=0,\,\mathcal{C}_q^{\prime}(\lambda)=0,\,\,\text{and}\,\,\mathcal{C}_q^{\prime\prime}(\lambda)=0$.
Solving these three constraints simultaneously in the parameter space
of $\alpha,\,\theta$ and $q$, we get two lines of third-order degeneracies,
which are given by

\begin{equation}
\theta=\theta_{1}(\alpha),\quad q=q_{1}(\alpha),\label{eq:3-EP-conditions-1}
\end{equation}
and

\begin{equation}
\theta=\theta_{2}(\alpha)=\pi-\theta_{1}(\alpha),\quad q=q_{1}(\alpha),\label{eq:3-EP-conditions-2}
\end{equation}
where
\begin{align}
\theta_{1}(\alpha) & =\frac{1}{2}\arccos\left(\frac{\alpha^{4}-8\alpha^{2}+1}{6\alpha^{2}}\right),\\
q_{1}(\alpha) & =-\frac{8\sqrt{6}\alpha\left(\alpha^{2}-1\right)^{3/2}}{3\left(\alpha^{4}-14\alpha^{2}+1\right)}.
\end{align}

The physical restriction on the postselection parameter, $0\le q\le1$,
implies that $\alpha\in[1,\sqrt{3}]$ in the above formulas. For $\alpha=1$,
we have $\theta_{1}(\alpha)=\theta_{2}(\alpha)=\pi/2$ and $q_{1}(\alpha)=0$,
i.e., the 3rd order degeneracy lines meet at the 4th order degeneracy,
cf.~Eq.~(\ref{eq:4-deg-conditions}).

Performing Jordan decomposition of $\mathcal{L}_{q}$ at the parameters
corresponding to the 3rd order degeneracies, one finds that these
are genuine 3rd order EPs.

\subsection{Second-order degeneracies\label{subsec:Second-order-EPs}}

The 2nd order degeneracies require $\mathcal{C}_q(\lambda)=0$ and $\mathcal{C}_q^{\prime}(\lambda)=0$.
Solving these two constraints simultaneously in the parameter space
of $\alpha,\,\theta$ and $q$, we get the surfaces parametrized as

\begin{equation}
q_{1}(\alpha,\theta)=\frac{\sqrt{2\eta}\left(3\left(\alpha^{2}-1\right)-\eta\right)}{3\sqrt{3}\alpha\,\sin^{2}\theta},\label{eq:2-EP-condition-1}
\end{equation}
and

\begin{equation}
q_{2}(\alpha,\theta)=\sqrt{\frac{2}{3}}\frac{\sqrt{\left(2\left(\alpha^{2}-1\right)-\eta\right)}\left(\alpha^{2}-1+\eta\right)}{3\alpha\,\sin^{2}\theta},\label{eq:2-EP-condition-2}
\end{equation}
where $\eta=\left(\alpha^{2}-1+\sqrt{\left(\alpha^{2}-1\right)^{2}-12\alpha^{2}\cos^{2}\theta}\right)$.
The surface parametrized by $q_{1}(\alpha,\theta)$ is shown in purple
in Fig.~1(b); the two orange surfaces in
the figure are parametrized by $q_{2}(\alpha,\theta)$.

Checking the number of linearly independent eigenvectors corresponding
to the degenerate eigenvalue at each of the surfaces reveals that
the line $q=0$, $\theta=\pi/2$, $\alpha>1$ (i.e., $2\left(\alpha^{2}-1\right)-\eta=0$)
is a trivial 2nd order degeneracy having two eigenvectors, while all
the other points on the surfaces are genuine 2nd order EPs (they have
only one eigenvector for the degenerate eigenvalue).
\end{document}